\documentclass[runningheads]{llncs}
\usepackage{amssymb,amsmath}
\usepackage{pstricks}
\usepackage{pst-node}
\usepackage{epsfig}
\usepackage{latexsym}
\begin{document}
\pagestyle{headings}

\newcommand{\lb}{[\![} %

\newcommand{\rb}{]\!]} %

 \newenvironment{ex}{\begin{example}\rm\ }{\end{example}}

 \newenvironment{df}{\begin{definition}\rm\ }{\end{definition}}

 \newenvironment{pr}{\begin{proposition}\rm\ }{\end{proposition}}

\title{Defining Rough Sets by\\ Extended Logic Programs
\thanks{Originally published in proc. PCL 2002, a FLoC workshop;
eds. Hendrik Decker, Dina Goldin, J{\o}rgen Villadsen, Toshiharu Waragai
({\tt http://floc02.diku.dk/PCL/}).}
}

\author{Jan Ma\l uszy\'nski\inst{1} \and Aida Vit\'{o}ria\inst{2}}

\institute{Dept. of Computer and Information Science,\\ Link\"oping
           University, S 581 83 Link\"oping, Sweden\\ \email{janma@ida.liu.se}
           \and 
           Dept. of Science and Technology,\\ Link\"oping University, S
           601 74 Norrk\"oping, Sweden\\ \email{ aidvi@itn.liu.se}}

\maketitle

\begin{abstract}
  We show how {\it definite extended logic programs} can be used for
  defining and reason with rough sets. Moreover, a rough-set-specific
  query language is presented and an answering algorithm is outlined.
  Thus, we not only show a possible application of a paraconsistent
  logic to the field of rough sets as we also establish a link between
  rough set theory and logic programming, making possible transfer of
  expertise between both fields.
\end{abstract}

\section{Introduction}
\label{sec:intro}

This paper shows how the formalism of {\em rough sets}
\cite{paw82,paw91,kpps99} used for processing of uncertain and
contradictory data relates to paraconsistent logic programming
\cite{cp98}. This gives a basis for efficient implementation of
rough sets in logic programming.

Since mid-eighties rough sets have been a subject of intensive
research. The literature on rough sets includes both theoretical
studies and reports on applications (for more information and
bibliography see the home page of the Rough Set Society
\verb|http://www.roughsets.org|).

A rough set is usually defined by a decision table which can be seen
as a finite collection of ground positive and negative datalog facts.
The table may include both a positive fact and its negation, thus it
may represent inconsistent information.  The intuition is that a
decision table defines a set, and the inconsistent facts identify
elements with uncertain membership.  A table may also include multiple
occurrences of facts.  This makes it possible to introduce a
quantitative measure for membership.  One of the main concepts is that
of the {\em lower approximation} of a rough set, corresponding to the
elements that the decision table asserts as being only positive
examples.

Viewing rows of a decision table as datalog facts gives a basis for
extending rough sets to Rough Datalog.  In our previous work
\cite{am02}, we proposed such an extension. Rough datalog makes it
possible to define rough sets not only explicitly as collections of
facts (as the decision tables do) but also implicitly by rules.  The
fixpoint semantics of rough datalog links the predicates of a program
to rough relations.  However, having predicates denoting rough
relations rather than relations may cause some difficulty in
understanding the rules.  Furthermore, the intuition of a rule as a
definition of a rough set is quite complex, since it has to define
both positive facts and negative facts of the defined rough set.
Finally, compilation of datalog rules to Prolog, described in
\cite{am02}, may cause explosion of the number of Prolog clauses
necessary to deal with negative facts.

In this paper we propose a simplified approach based on the concept of
{\em definite extended logic program} (DXL programs) \cite{cp98}.  As
mentioned above, decision tables for rough sets include explicit
negative information. This information can be expressed in DXL
programs by using {\em explicit negation}. Thus, DXL programs are well
suited to represent rough sets. The fixpoint semantics of DXL
determines then the rough sets specified by a given program.  DXL
programs can be easily implemented and queried in pure Prolog.
However, DXL is not expressive enough for stating rough-set-specific
queries. For example, in DXL it is not possible to query lower
approximations of the defined rough sets.  To achieve this we propose
to extend DXL with a query language tailored for rough sets.  We also
show how to obtain answers by transforming the queries to usual Prolog
queries.

The rough sets denoted by the predicates occurring in a program are
similar to the paraconsistent relations used in \cite{rr96}.  The main
aim of the work presented in \cite{rr96} is to introduce an algebraic
method to construct the well-founded model \cite{grs91} for a general
deductive database by using paraconsistent relations associated with
each predicate symbol of the database.  Since the well-founded model
is always a consistent interpretation, predicates occurring in a
general deductive database denote crisp sets\footnote{Using rough set
  terminology, a crisp set has an empty boundary region.}.  However,
in contrast to \cite{rr96}, the models of the programs proposed in our
framework may incorporate contradictions.  Consequently, while
well-founded models are $3$-valued, we use models in a
\mbox{$4$-valued logic}. Moreover, we deal with explicit negation.

The rest of the paper is organized as follows.  Section \ref{sec:rs}
surveys some basic concepts of rough sets.  Section \ref{sec:dxl}
summarizes the semantics of DXL programs and gives an example of how
rough sets can be represented via DXL programs.  Section
\ref{sec:query} discusses a rough-set-specific query language and
proposes an algorithm to obtain answers.  Section \ref{sec:concl}
gives some conclusions.

\section{Rough sets}
\label{sec:rs}

This section gives a brief introduction to Rough Sets. 

We want to deal with the situation where there are conflicting
judgments about classification of a given object.  For example, two
patients show identical results of clinical tests but one of them has
a certain disease and the other does not have it, or the experts
looking at the medical record of a patient may disagree on the
diagnosis. The concept of rough set makes possible to express such a
situation. More precisely, the situation can be described as follows.
We have a universe of objects, each of them characterized by a tuple
of attribute values and by a decision attribute classifying the
object. For simplicity, we assume two-valued (say ``yes'' or ``no'')
classification.  This can be seen as a definition of a set consisting
of all objects with the decision attribute ``yes''. However, objects
with identical attribute values may have different values of the
decision attribute.  Since we only access objects by their attribute
values, the double classification describes the boundary region, where
we cannot be sure whether the object belongs to the defined set or
not.  Thus intuitively, a rough set $S$ is defined by indicating the
elements of a universe which belong to $S$ and elements which do not
belong $S$, while these two categories need not be disjoint.  Usually,
it is also assumed that the union of these categories covers the
universe. In practice this may be achieved by the assumption that the
elements which do not appear in the decision table are implicitly
classified as not belonging to the defined set.  In this paper we do
not make this assumption. This makes it possible to distinguish
between the tuples whose membership in $S$ is explicitly negated and
those for which we have no membership evidence. This distinction is
well known in the field of logic programming, while it seems not to be
discussed in the context of rough sets.  Notice that our assumption
does not exclude the possibility that the union of both categories
covers the universe, and thus generalizes the usual approach.

\begin{example}
\label{rset}
The following table contains patient records with the symptom
attributes {\em temp}erature, {\em cough}, {\em headache}, {\em
  muscle-pain} and the diagnosis done by a doctor which says whether
or not the patient has flu. The table defines a rough set, since it
includes different diagnoses for some cases with identical symptoms.

\begin{center}
{\small
\begin{tabular}{|l|l|l|l||l||} \hline
{\em temp} & {\em cough} & {\em  headache} & {\em muscle pain} & {\bf flu} \\ \hline \hline
normal & no & no & no & no \\ \hline
subfev & no & yes & yes & no \\ \hline
subfev  & no & yes & yes & yes \\ \hline
subfev & yes & no  & no & no \\ \hline
subfev & yes & no & no & yes \\ \hline
high & no  & no & no & no \\ \hline 
high & yes & no & no & no \\ \hline 
high & yes & no & no & yes \\ \hline
high & yes & yes & yes  & yes \\ \hline
\end{tabular}
}
\end{center} 
\end{example}

\vspace{5mm}
The intuitions discussed  above can be formalized by
the following definitions.

An attribute $a$ is  a  function $a:U
\rightarrow V_a$, where $U$ is a universe of objects. The set $V_a$ is
called the {\em value domain} of $a$. 

We assume that tuples of values provide the only way of referring to
objects.  Two objects are \emph{indiscernible} with respect to a
selected set of attributes, if both have the same values for these
attributes. Clearly, the indiscernibility relation is an equivalence
on objects and its equivalence classes are sets of objects which are
characterized by identical tuples of attribute values. We assume that
the tuples provide the only access to objects. Hence, the technical
definitions are expressed in terms of tuples.  As illustrated by the
table above, we specify a rough set $S$ by classifying tuples of
attribute values as positive or negative examples.

\begin{df}
  A {\em rough set} $S$ is a pair $(S^+,S^-)$ such that $S^+, S^-
  \subseteq V_{a_1} \times \cdots \times V_{a_n}$, for some non empty
  set of attributes $\{a_1, \cdots, a_n\}$.
\end{df}

The components $S^+$ and $S^-$ will be called the \emph{positive
  region} (or the \emph{positive information}) and the \emph{negative
  region} (or the \emph{negative information}) of $S$, respectively.

We will also use the following notion.

\begin{df}
\label{df:comp}
A {\em rough complement} of a rough set $S = (S^+,S^-)$ is the rough
set $\neg S = (S^-,S^+)$.
\end{df}

Using rough set terminology, given a rough set $S = (S^+, S^-)$, the
sets $S^+$ and $(S^+ - S^-)$ correspond to the \emph{upper
  approximation} and the \emph{lower approximation} of $S$,
respectively. Thus, the approximations of $\neg S$ are:
 $S^-$ (the upper approximation) and  $(S^- - S^+)$ %
 (the lower approximation). The set $S^+ \bigcap S^-$ is called the
 {\em boundary (region)} of $S$. Intuitively, the lower approximation
 of $S$ ($\neg S$) refers to the elements that can certainly be
 classified as (not) members of $S$. The elements in the boundary may
 belong to $S$ ($\neg S$), but we cannot be sure. It is easy to see
 that both the lower approximation and boundary of $S$ are subsets of
 the upper approximation of $S$.

 The following definition, adopted from \cite{kpps99} formalizes the
 idea of the decision system (also called decision table) used to
 define rough sets.

\begin{df}
  A {\em (binary) decision system} is a pair ${\cal D} =(U,A\cup
  \{d\})$, where $U$ is a universe of objects, and $A \cup \{d\}$ is a
  non-empty finite set of {\em attributes}, such that $d:U \rightarrow
  \{true,false\}$. We allow that for some $u\in U$ all attribute
  values, including the value of $d$, are undefined.
\end{df}

For a given $u\in U$ and set of attributes $A = \{a_1, \cdots, a_n\}$,
we denote by $A(u)$ the tuple $\langle a_1(u), \cdots, a_n(u) \rangle$.
Recall that $A$ may be undefined for some $u$. Thus, $A$ is a partial
function on objects.

\begin{df}
\label{dt}
  {\em A rough set $D$ specified by a decision system ${\cal D} =
    (U,A\cup\{d\})$} is a pair $(D^+, D^-)$, where 
\begin{center}
  $D^+ = \{A(u)|\; u \in U \mbox{ and } d(u) = \mbox{true}\}$,\\
  $D^- = \{A(u)|\; u \in U \mbox{ and } d(u) = \mbox{false}\}$.
\end{center}
\end{df}

\begin{example}
\label{rset-cont}
Consider the rough set \textsf{Flu} specified by the decision system
of Example \ref{rset}.

It is easy to check that the lower approximation of \textsf{Flu} is
the singleton $\{\langle\mbox{high, yes, yes, yes}\rangle\}$. The set
$\{\langle\mbox{normal, no, no, no}\rangle, \langle\mbox{high, no, no,
  no}\rangle\}$ is the lower approximation of the rough set
$\neg$\textsf{Flu}.  The boundary region of \textsf{Flu} consists of
all other remaining tuples in the decision table.
\end{example}

A binary decision system can be equivalently represented by a set of
literals. We illustrate the idea on the decision table of Example
\ref{rset}. We assume \verb|flu| to be a 4-ary predicate letter. Each
row of the table is then represented by a literal with the argument
values stated in the row. The literal is positive if the decision
attribute's value is ``yes'' and negative otherwise.  Thus, we obtain
the set of literals:
\[
\begin{array}{l}
\{\neg \verb|flu(normal,no,no,no)|, \neg\verb|flu(subfev,no,yes,yes)|,\\ 
  \hspace{2mm}\verb|flu(subfev,no,yes,yes)|, \cdots\}\; .
\end{array}
\]

\section{ Definite Extended Logic  Programs}
\label{sec:dxl}

This section recalls the concept of Definite Extended Logic Programs
and relates them to rough sets. Definite extended logic programs
extend classical definite logic programs with explicit negation.
Similar ideas were discussed by many authors, see e.g.
\cite{bs89,pw91,w93,w94}. We follow here the presentation of the
survey paper \cite{cp98}.

As discussed above, a rough set $S$ can be defined by providing
explicitly a set of literals with the same predicate letter.  The
positive literals (e.g. $s(t_1, \cdots, t_n)$) identify the tuples in
the positive region of $S$, while the negative literals (e.g. $\neg
s(t_1, \cdots, t_n)$) determine its negative region.  This can be seen
as an alternative representation of a decision system.

Definite Extended Logic Programs provide a more general way 
of defining sets of literals. 

\begin{df}\cite{cp98}
A {\em definite extended logic program (DXL program)}  
is a set of rules of the form
\[H :-\,  B_1, \cdots ,B_n.\mbox{\hspace{1cm}}(n\geq 0)\]
where $H,B_1, \cdots,B_n$ are literals.
\end{df}

Notice that rules extend definite clauses by allowing negative
literals, both in the head and in the body.  In the sequel, the rules
with empty bodies (facts) will be written in the form $H.\, \,$.

The semantics of DXL programs is defined by viewing each negated
literal $\neg p(t_1, \cdots, t_n)$ as a positive literal $p^-(t_1,
\cdots, t_n)$, with a new predicate symbol $p^-$. In this way, a DXL
program ${\cal P}$ is transformed into a definite program ${\cal P}'$.
The standard least Herbrand model semantics ${\cal M}_{{\cal P}'}$ of
${\cal P}'$ is a set of ground atoms, over the original and the new
predicate symbols.  The semantics of the DXL program ${\cal P}$,
${\cal M}_{\cal P}$, is defined by replacing each atom of the form
$p^-(t_1, \cdots, t_n) \in {\cal M}_{{\cal P}'}$ by the corresponding
negative literal $\neg p(t_1, \cdots, t_n)$.

Clearly, in general ${\cal M}_{\cal P}$ may include an atom together
with its negation. Thus, a DXL program ${\cal P}$ may introduce
inconsistencies.  This is what is needed to be able to define rough
sets. Each predicate symbol $p$, with arity $n \geq 0$, occurring in
${\cal P}$ denotes the rough relation (set)
$$\mbox{\textsf{P}} = (\{(t_1, \cdots, t_n)\, |\,p(t_1, \cdots, t_n)
\in M_{\cal P}\}, \{(t_1, \cdots, t_n)\,|\,\neg p(t_1, \cdots, t_n)
\in M_{\cal P}\} )\,\, .$$

For a  model theoretic semantics for DXL programs based on the four-valued
Belnap's logic the reader is referred to \cite{b77}.

We now show an example of a definition of rough sets by a DXL program.

\begin{ex}
\label{epat}
We consider the rough relation \textsf{Flu} of Example \ref{rset} and
a rough relation \textsf{Patient} with the same attributes as
\textsf{Flu} extended with the new ones: {\em id}entification, {\em
  age} and {\em sex}.  Intuitively, the universe of relation
\textsf{Patient} is a set of people who visited a doctor. Its decision
attribute shows whether a person has to be treated for some disease
and, therefore, has to be considered a patient.  The decision may be
made independently by more than one expert. All decisions are
recorded, what might make the relation rough.  The example relation is
defined by the following decision table.

\vspace{0.5cm}
\begin{center}
{\small
\begin{tabular}{|l|l|l|l|l|l|l||l||} \hline
{\em id} & {\em age} & {\em sex}&
{\em temp} & {\em cough} & {\em  headache} & {\em muscle pain} & {\bf patient} \\ \hline \hline
1 & 21 & m & normal & no & no & no & no \\ \hline
2 & 51 & m&  subfev & no & yes & yes & yes \\ \hline
3  & 18  & f&  subfev & no & yes & yes   & no \\ \hline
3  & 18  & f&  subfev & no & yes & yes   & yes \\ \hline
4  & 18  & m&  high  & yes & yes & yes  & yes \\ \hline
\end{tabular} 
}
\end{center}

\vspace{0.5cm} In order to know who are the people to be treated for
flu, we define a new rough set \textsf{Ft}.  Intuitively, these are
people possibly qualified as patients, who may have flu according to
the decision table of Example \ref{rset}. We may also state that the
people not treated for flu are those not qualified as patients; or
those qualified as patients who may not have flu.

This can be expressed as the following DXL program ${\cal P}$.

\vspace{5mm}
\noindent
\verb|ft(Id):- patient(Id,Age,Sex,Fev,C,Ha,Mp),| \\
\verb|         flu(Fev,C,Ha,Mp).|\\
$\neg$\verb|ft(Id):- |$\neg$\verb|patient(Id,Age,Sex,Fev,C,Ha,Mp).|\\
$\neg$\verb|ft(Id):- patient(Id,Age,Sex,Fev,C,Ha,Mp),|\\ 
\verb|          |$\neg$\verb|flu(Fev,C,Ha,Mp).|
\vspace{5mm}

As explained above, the semantics of this program determines the rough
relation \textsf{Ft}. Thus, we can conclude that person $4$ is
definitely qualified for flu treatment (i.e. belongs to the lower
approximation of \textsf{Ft}).  Persons $2$ and $3$ may or may not be
treated for flu (i.e. belong to the boundary of \textsf{Ft}) and
person 1 is not certainly qualified for flue treatment (i.e. belongs
to the lower approximation of $\neg$\textsf{Ft}).
\end{ex}

\section{Rough  Set Queries}
\label{sec:query}

The transformed version ${\cal P}'$ of a DXL program ${\cal P}$,
defined above, may be used by a Prolog system for answering queries
about rough sets.  Notwithstanding the incompleteness of Prolog we
conclude, that whenever the query evaluation terminates and succeeds,
we obtain an answer showing an instance of the query consisting of the
elements of the least model.

Other systems exist that can answer queries w.r.t to a normal program,
for instance, {\it XSB}-Prolog (for more details see
\verb|http://xsb.sourceforge.net/|). Hence, also those systems could
be used to implement our query answering algorithm.

\subsection{A Query Language for Rough Sets}
\label{subsec:queries}

Since the proposed query answering technique refers to the least model
of the transformed program, in the terminology of rough sets the
answer concerns the upper approximations of the defined rough sets.
For example, consider program ${\cal P}$ of Example \ref{epat}, the
answer \verb|yes| to the query\, \verb|?|(${\cal P}$, \verb|ft(4)|)\,
means that person $4$ belongs to the upper approximation of the rough
set \textsf{Ft}.  However, it may also be important to check whether a
given element is in the lower approximation of a rough set, or what
are the elements in the boundary region of a given set. Thus, we
propose to extend DXL with the following rough set specific queries.

\begin{df}
\label{roughQuery}
A {\em rough query} ${\cal Q}$ is a pair\, \verb|?|$({\cal P}, q)$\,,
where ${\cal P}$ is a DXL program and $q$ is defined by the following
abstract syntax rules
$$\begin{array}{lll} 
   q & \longrightarrow & q' \mid a? \\
   q' & \longrightarrow & l \mid \underline{l} \mid
        \overline{\underline{a}} \mid  q'_1, q'_2\,\, ,
\end{array}$$
where $l$ is a literal and $a$ is an atom.

\end{df}

Let ${\cal Q} = \mbox{\texttt{?}}({\cal P}, q)$ be a query, given a
DXL program ${\cal P}$. Then, ${\cal Q}$ is a {\em simple query} if
$q$ is a literal $l$, or of the form $\underline{l}$ or
$\overline{\underline{a}}$, where $a$ is an atom. A {\em composite
  query} is a sequence of simple queries, separated by commas. A
composite query is interpreted as a conjunction of simple queries.

Let ${\cal P}$ be a DXL program and \textsf{R} be the rough relation
denoted by predicate $r$ of ${\cal P}$. First, we explain intuitively
how the answer to a ground simple query can be obtained.  The answer
to a ground simple query may only be \verb|yes| or \verb|no|.
\begin{itemize}
\item The answer to a query\, \verb|?|$({\cal P},r(t_1, \cdots,
  t_n))$\, (\,\verb|?|$({\cal P},\neg r(t_1, \cdots, t_n))$\,) is
  \verb|yes| {\it iff} the tuple $(t_1, \cdots, t_n)$ belongs to the
  positive region (negative region) of the rough relation \textsf{R},
  defined by ${\cal P}$. Otherwise, the answer is \verb|no|.
\item The answer to a query\, \verb|?|$({\cal P},\underline{r}(t_1,
  \cdots, t_n))$ (\,\verb|?|$({\cal P},\underline{\neg r}(t_1, \cdots,
  t_n))$\,) is \verb|yes| {\it iff} the tuple $(t_1, \cdots, t_n)$
  belongs to the lower approximation of \textsf{R} ($\neg$\textsf{R}).
  Otherwise, the answer is \verb|no|.
\item The answer to a query\, \verb|?|$({\cal
    P},\overline{\underline{r}}(t_1, \cdots, t_n))$\, is \verb|yes|
  iff the tuple $(t_1, \cdots, t_n)$ belongs to the boundary region of
  \textsf{R}. Otherwise, the answer is \verb|no|.
\end{itemize}

The ground query\, \verb|?|$({\cal P}, r(t_1, \cdots, t_n)?)$\,
questions what is known about atom $r(t_1, \cdots, t_n)$ in the least
model of ${\cal P}$. Four cases are possible.  Tuple $(t_1, \cdots,
t_n)$ may belong to the boundary region of the denoted rough set
\textsf{R}, to its lower approximation, to the lower approximation of
$\neg$\textsf{R}, or to none of these.  The respective answers will
be: $\top$, \verb|yes|, \verb|no|, and $\bot$. Notice that $\top$
represents the existence of contradictory information and $\bot$
represents absence of information. Although this kind of queries are
not strictly needed because the same information can be obtained with
several simple queries, they might be useful in practice. For
instance, for a given $n$-ary predicate $r$, the query \,
\verb|?|$({\cal P}, r(X_1, \cdots, X_n)?)$\, classifies all possible
$n$-ary tuples with respect to the membership of the rough relation
denoted by $r$.

A natural extension to non-ground simple queries\, \verb|?|$({\cal P},
q)$\, case (i.e. $q$ contains some variables) gives as answer the set of
all valuations $\theta$ for which the query instance\, \verb|?|$({\cal
  P},\theta(q))$\, satisfies the above mentioned conditions. The answer
\verb|no| represents the empty set of valuations, and the answer
\verb|yes| corresponds to the set of all ground valuations of the
variables of the query.

The answer to a query of the form\, \verb|?|$({\cal P}, r(t_1, \cdots,
t_n)?)$\,, where $r(t_1, \cdots, t_n)$ is a non ground atom, is a
triple of sets $(A_1, A_2, A_3)$: set $A_1$ corresponds to the
instances of the query that belong to the boundary of \textsf{R}; set
$A_2$ corresponds to the instances of the query that belong to the
lower approximation of \textsf{R}; set $A_3$ corresponds to the
instances of the query that belong to the lower approximation of
$\neg$\textsf{R}. Obviously, answers to this type of queries can be
obtained by issuing the simple queries \verb|?|$({\cal P},
\overline{\underline{r}}(t_1, \cdots, t_n))$\, , \verb|?|$({\cal P},
\underline{r}(t_1, \cdots, t_n))$\, and \verb|?|$({\cal P},
\underline{\neg r}(t_1, \cdots, t_n))$\,.

The above ideas can be easily extended to the case of composite
queries. Note that a query of the form\, \verb|?|$({\cal P}, q?)$\,
cannot be involved in a composite query.

\begin{example}
  Consider the program ${\cal P}$ of Example \ref{epat}, defining the
  rough relation \textsf{Ft}. We may pose queries like\,
  \verb|?|$({\cal P},\mbox{\texttt{ft(3)}})$\,, \verb|?|$({\cal
    P},\overline{\underline{\mbox{\texttt{ft}}}}\mbox{\texttt{(X)}})$\,
  or\, \verb|?|$({\cal P},\neg\mbox{\texttt{ft(4)}})$\,.  The obtained
  answer would then be: \verb|yes| for the first query;
  $\{$\verb|X=2,X=3|$\}$ for the second one; and \verb|no| for the
  last one.
\end{example}

\subsection{Implementing Rough Queries in Prolog}

As already discussed the simple literal queries for a DXL program
${\cal P}$ can be directly answered in Prolog by using the transformed
version ${\cal P}'$ of ${\cal P}$.

We now show how the remaining queries can also be answered by
transforming them to Prolog queries for $P'$. We define the following
transformation $\tau$ of simple queries to Prolog queries, where {\it
  not} denotes Prolog negation as failure (\verb|\+|).

\[ \tau(Q) = \left \{ \begin{array}{l@{\hspace{4mm}}l}
                     q(t_1, \cdots, t_n)   & \mbox{if $Q \equiv q(t_1, \cdots, t_n)$}\\
                     q^-(t_1, \cdots, t_n)   & \mbox{if $Q \equiv \neg q(t_1, \cdots, t_n)$}\\
                     q(t_1, \cdots, t_n), \mbox{{\it not} }q^-(t_1, \cdots, t_n)
                            & \mbox{if $Q \equiv \underline{q}(t_1, \cdots, t_n)$}\\
                     q^-(t_1, \cdots, t_n), \mbox{{\it not} }q(t_1, \cdots, t_n)
                            & \mbox{if $Q \equiv \underline{\neg q}(t_1, \cdots, t_n)$}\\
                     q(t_1, \cdots, t_n), q^-(t_1, \cdots, t_n)
                            & \mbox{if $Q \equiv \overline{\underline{q}}(t_1, \cdots, t_n)$}
                     \end{array}
              \right. \]

              Let ${\cal P}$ be a DXL program. We now claim that the
              answers obtained by Prolog evaluation of the query
              $\tau(Q)$ w.r.t to the program ${\cal P}'$ coincide with
              the answers defined for\, \verb|?|$({\cal P},Q)$\,, in
              Section \ref{subsec:queries}.  Let $p$ be a predicate
              letter occurring in ${\cal P}$.  By the construction of
              ${\cal P}'$, it follows that an atom $p(t_1, \cdots,
              t_n)$ belongs to ${\cal M}_{{\cal P}}$ iff it also
              belongs to ${\cal M}_{{\cal P}'}$. Moreover, a negative
              literal $\neg p(t_1, \cdots, t_n) \in {\cal M}_{{\cal
                  P}}$ iff the atom $p^-(t_1, \cdots, t_n) \in {\cal
                M}_{{\cal P}'}$.  Recall that ${\cal P}'$ is a
              definite program. If a simple query $q(t_1, \cdots,
              t_n)$ or $q^-(t_1, \cdots, t_n)$ w.r.t ${\cal P}'$ fails
              in Prolog, then it has no (ground) instances in ${\cal
                M}_{{\cal P}'}$. In view of that, it can be easily
              checked that each of the five cases of the definition of
              ${\tau}$ satisfy our claim. Take for example a lower
              approximation rough query\, \verb|?|$({\cal P},
              \underline{q}(t_1, \cdots, t_n))$\,.  Assume that the
              Prolog answer to $\tau(\underline{q}(t_1, \cdots, t_n))$
              w.r.t ${\cal P}'$ returns a valuation $\theta$. Thus,
              $\theta(q(t_1,\cdots,t_n))$ is in ${\cal M}_{{\cal
                  P}'}$, hence in ${\cal M}_{{\cal P}}$. On the other
              hand, if $\theta(q^-(t_1,\cdots,t_n))$ fails then
              $\theta(q^-(t_1,\cdots,t_n))\not\in {\cal M}_{{\cal
                  P}'}$. Thus, $\theta( \neg q(t_1,\cdots,t_n))$ is
              not in ${\cal M}_{{\cal P}}$. Consequently, $\theta(
              q(t_1,\cdots,t_n))$ belongs to the lower approximation
              of the rough set \textsf{Q} denoted by predicate $q$, as
              required. One should also consider the case when the
              Prolog query $\tau(Q)$ fails w.r.t ${\cal P}'$. This
              means that $q(t_1,\cdots,t_n)$ fails (i.e. there is no
              instance of $q(t_1,\cdots,t_n)$ that belongs to the
              upper approximation of \textsf{Q}) or that whenever
              $q(t_1,\cdots,t_n)$ succeeds with a valuation $\theta$
              then $\mbox{{\it not} } \theta(q^-(t_1,\cdots,t_n))$
              fails (i.e. $\theta(\neg q(t_1,\cdots,t_n))$ is in the
              negative region of \textsf{Q}).  Thus, in both cases
              there is no instance of the query which belongs to the
              lower approximation of the rough relation \textsf{Q}.

\section{Discussion and Conclusions}
\label{sec:concl}

The contribution of the paper is twofold. First, it establishes a link
between logic programming and rough set theory that makes possible to
combine techniques originating from both fields. Second, we show an
application of the techniques developed in the area of paraconsistent
logic.

We relate DXL programs to rough sets: we have shown that the least
model of any DXL program can be seen as a family of rough relations.
Although this observation is technically very straightforward, it
opens for use of Prolog for defining and manipulation of rough sets.
To our knowledge this approach is novel as concerns rough sets. It
improves and simplifies our recent work on rough datalog \cite{am02}, by
providing more flexible technique for defining negative regions of
rough sets, which results in simplification of the semantics.

The language of rough queries brings the specificity of rough sets to
paraconsistent logic programming. It should be clear that with this
language, mainly due to the use of lower approximations, we implicitly
introduce a very restricted form of default negation into DXL. A
natural question is whether the lower approximations should be
introduced into bodies. There may be example applications such that
the reference to lower approximations in the rules may be desirable.
However, so far the interest of rough sets community for nonmonotonic
reasoning seems to be rather limited.

Extension of the language with lower approximations in the body would
require a more sophisticated semantics. However, such an extension
would still not allow a free use of default negation. An interesting
question is then whether these restrictions make it possible to
provide a simple and intuitive semantics.

\newpage

\end{document}